\documentclass[runningheads]{llncs}
\usepackage{cite}
\usepackage{amsmath,amssymb,amsfonts}
\usepackage{algorithmic}
\usepackage{graphicx}
\usepackage{textcomp}
\usepackage{xcolor}
\usepackage{multirow}
\usepackage{listings}
\usepackage{rotating}
\usepackage{subcaption}
\usepackage{hyperref}

\begin{document}
\title{DeepClone: Modeling Clones to Generate Code Predictions\thanks{The final authenticated version is available online at \url{https://doi.org/10.1007/978-3-030-64694-3_9}}}

\author{Muhammad Hammad\inst{1}\thanks{Corresponding Author}\and
\"{O}nder Babur\inst{1} \and
Hamid Abdul Basit\inst{2} \and
Mark van den Brand \inst{1}}

\authorrunning{Hammad et al.}


\institute{Eindhoven University of Technology, Netherlands
\email{\{m.hammad,o.babur,m.g.j.v.d.brand\}@tue.nl}\and
Prince Sultan University, Saudi Arabia\\
\email{hbasit@psu.edu.sa}}

\maketitle              
\begin{abstract}
Programmers often reuse code from source code repositories to reduce the development effort. Code clones are candidates for reuse in exploratory or rapid development, as they represent often repeated functionality in software systems. To facilitate code clone reuse, we propose \emph{DeepClone}, a novel approach utilizing a deep learning algorithm for modeling code clones to predict the next set of tokens (possibly a complete clone method body) based on the code written so far. The predicted tokens require minimal customization to fit the context. DeepClone applies natural language processing techniques to learn from a large code corpus, and generates code tokens using the model learned. We have quantitatively evaluated our solution to assess (1) our model's quality and its accuracy in token prediction, and (2) its performance and effectiveness in clone method prediction. We also discuss various application scenarios for our approach. 


\end{abstract}

\section{Introduction}
Writing new code is an expensive activity, consuming considerable time and effort. Developers frequently perform adhoc code reuse, searching for code snippets over the web or in some codebase, followed by judicious copying and pasting\cite{gharehyazie2017some}. Features like code snippet search, code prediction, code auto-completion and code generation can help developers to write code quickly and easily. Lately, language modeling have been effectively employed for these tasks\cite{radford2019language,karampatsis2020big,zhong2019javascript}.

A Language Model (LM) estimates the likelihood of sequences of tokens based on a training dataset, by assigning probabilities to tokens (words, subwords, or punctuation marks) or character sequences (sentences or words occurring after a given sequence \cite{karampatsis2020big}). Shannon first used language modeling \cite{shannon1951prediction} to predict the next element following some given English text. Since then, several language models have been developed to perform different tasks. Various statistical and Deep Neural Networks (DNN) based techniques for language modeling have been applied to natural languages. These techniques are also applicable to programming languages \cite{allamanis2013mining,boldt2017using,hellendoorn2017deep,white2016deep}. 

DNN techniques are powerful machine learning models that perform well in language modeling for source code, outperforming statistical language modeling techniques \cite{karampatsis2020big}. Performance of DNN techniques improves automatically through experience by learning data patterns; their power arising from the ability to perform parallel computations for a large number of training steps. LMs constructed using DNNs are called Neural Language Models (NLM), which have been used for various software development tasks like code completion\cite{karampatsis2020big,zhong2019javascript} and code clone detection \cite{white2016deep} etc. 

One common application of language modeling is code prediction \cite{allamanis2013mining,boldt2017using} - a technique for predicting next likely  token(s) on the basis of user input (i.e., code written so far), by learning the source code features. These code predictions have a fixed threshold for the length of generated token sequence. In this work, we show how clone methods of arbitrary length, extracted from a large code repository, can enhance regular code generation and prediction applications. 

Code clones are repeated patterns in code, usually created with the copy-paste ad hoc reuse practice. Around 5 to 50\% of the code in software applications can be contained in clones \cite{roy2007survey}. Clones are generally considered harmful, and several techniques exists for avoiding and eliminating them \cite{hammad2020systematic}. However, clones can be useful in certain software development scenarios \cite{kapser2008cloning}, one of them being exploratory development where the rapid development of a feature is required, and the remedial unification of newly generated clone is not clearly justified. Also, a piece of cloned code is expected to be more stable and poses less risk than new development. 

We believe that clone methods, together with the non-cloned code, can be a useful component of a LM, as they represent commonly used functionality, and can be used for code prediction and completion. In this work, we exploit the re-usability aspect of code clones to build a NLM for predicting code tokens up-to the level of method granularity. We believe that our approach can help in improving the quality of code prediction by also predicting complete method bodies of arbitrary length based on clone methods. In this work, we have made the following contributions:

\begin{enumerate}
 \item We present a novel approach for code prediction by explicitly modeling code clones along with non-cloned code.
 \item Our approach can generate code predictions of complete method body of arbitrary length on the basis of user input.
 \item We have quantitatively evaluated our approach using the BigCloneBench dataset, in terms of model quality and performance in various tasks including token prediction, and clone method prediction.
\end{enumerate}

\section{Related Work}
To the best of our knowledge, no previous techniques has modeled code clones together with non-cloned code for predicting code tokens up-to the complete method granularity. However, many techniques have explored language modeling for token prediction, code suggestions or code completion. White et al. \cite{white2015toward} applied Recurrent Neural Network (RNN) to model Java source code for code prediction. Bold \cite{boldt2017using} modeled Java language method statements and English language datasets by using Long Short Term Memory (LSTM). He compared the performance of next token prediction task with each other, arguing that method statements highly resemble English language sentences and are comparable to each other. Hellendoorn and Devanbu \cite{hellendoorn2017deep} noticed that source code NLMs under-perform due to the unlimited vocabulary size as new identifiers keep coming with higher rates, and limiting vocabulary is not a good solution for NLMs. They proposed a nested scope, dynamically updatable, unlimited vocabulary count-based n-gram model, which outperforms the LSTM model on the task of token prediction. In contrast, Karampatsis et al. \cite{karampatsis2020big} solved the issue of unlimited vocabulary size by applying byte-pair encoding (BPE) technique in modeling the code. They compared the performance of n-gram and Gated Recurrent Unit (GRU) language models trained on source code datasets, and demonstrated that NLM trained on GRU can outperform n-gram statistical models on code completion and bug detection tasks if BPE technique is applied. Zhong et al. \cite{zhong2019javascript} applied the LSTM model with sparse point network to build a language model for JavaScript code suggestion. Deep TabNine \footnote{\url{https://tabnine.com/blog/deep}} is a recently developed software programming productivity tool, successfully fine-tuned by using GPT-2 on approximately two million GitHub files capturing numerous programming languages, to predict the next chunk of code.

\section{BigCloneBench for Code Clones}
\label{sec:bigclonebench}
BigCloneBench\cite{svajlenko2014towards,svajlenko2016bigcloneeval} is the largest clone benchmark dataset, consisting of over 8 million manually validated clone method pairs in IJaDataset 2.0 \footnote{\url{https://sites.google.com/site/asegsecold/projects/seclone}}- a large Java repository of 2.3 million source files (365 MLOC) from 25,000 open-source projects. BigCloneBench contains references to clones with both syntactic and semantic similarities. It contains the references of starting and ending lines of method clones existing in the code repository. In forming this benchmark, methods that potentially implement a given common functionality were identified using pattern based heuristics. These methods were manually tagged as true or false positives of the target functionality by judges. All true positives of a functionality were grouped as a clone class, where a clone class of size $n$ contains $\frac{n(n-1)}{2}$ clone pairs. The clone types and similarity of these clone pairs were later identified in a post-processing step. Currently, BigCloneBench contains clones corresponding to 43 distinct functionalities. Further details can be found in the relevant publications \cite{svajlenko2014towards,svajlenko2016bigcloneeval}.

\subsection{Dataset Preparation}
\label{sec:datapreperation} 
We are using a reduced version of IJaDataset containing only the source files whose clone method references exist in BigCloneBench \cite{svajlenko2014towards,svajlenko2016bigcloneeval}. The dataset is distributed into a number of smaller subsets, on the basis of 43 distinct functionalities. 

We have performed several pre-processing steps to build our mutually exclusive training, testing, and validation datasets. First, we filtered IJaDataset files, keeping those which have references of true positive clone methods and discarding false positive clone references in BigCloneBench dataset (\textit{Filtering}). Next, we distributed the set of files into training, validation, and testing datasets (\textit{Distribution}). We adopted stratified sampling \cite{trost1986statistically} to ensure that all types of clone methods appear in training, validation, and testing datasets. We distributed the set of files in each functionality folder into portions as per the following ratio: 80\% training, 10\% validation, and 10\% testing. Then, we copied those files from original distribution to three separate folders such as training, validation, and testing. If any of the file already exist in one of those folders, we discarded it to avoid exact duplication \cite{allamanis2019adverse}. Tables \href{https://www.win.tue.nl/~mhammad/deepclone.html}{\textcolor{red}{A5}}\footnote{Tables A1, A2, A3, A4 and A5 can be accessible through link \url{https://www.win.tue.nl/~mhammad/deepclone.html}} 
and \ref{tab:experimentalcorpus} show the statistics of our datasets. Next, we marked the clone methods in the IJaDataset files by placing the meta-token \big \langle soc\big\rangle~at the start, and \big \langle eoc\big\rangle~at the end (\textit{Marking}). We normalized our code by removing whitespaces, extra lines, comments (\textit{Normalization}), and tokenized it by adapting Java 8 parser from Javalang\footnote{\url{https://github.com/c2nes/javalang}} Python library. We also replaced integer, float, binary, and hexadecimal constant values with the \big \langle num\_val\big \rangle~meta-token (\textit{Replacement}). Similarly, string and character values were replaced with \big \langle str\_val\big \rangle. This reduced our vocabulary size, leading to faster training of the model \cite{white2015toward,karampatsis2020big}. Finally, we merged all the tokenized data from the training, validation and testing files into three text files, i.e.~ train.txt, valid.txt, and test.txt (\textit{Merging}). 
Table \href{https://www.win.tue.nl/~mhammad/deepclone.html}{\textcolor{red}{A4}} demonstrates the pre-processing steps on an example of binary search clone method, while Table \ref{tab:experimentalcorpus} gives an overview of our experimental dataset.

\section{Neural Language Models for Code Clones}
A number of techniques were available for developing an LM for BigCloneBench dataset such as n-gram \cite{hellendoorn2017deep}, LSTM\cite{hochreiter1997long}, GRU\cite{cho2014learning},  GPT-2 \cite{radford2019language}; as well as parameter settings for training those models. We could not evaluate all the possible combinations (hundreds) and especially very large scale models/training due to the resource limitations. We selected GRU\cite{cho2014learning} and GPT-2\cite{radford2019language} as they have been reported to outperform other comparable models with recommended configurations. In the following sections we describe the two models.

\subsection{Gated Recurrent Units (GRU)}
Gated recurrent units (GRUs) are a gating mechanism in RNNs \cite{cho2014learning}, which is similar to LSTM \cite{hochreiter1997long} but has a forget gate and fewer parameters as it lacks an output gate. However, it is known to perform better than LSTM on certain tasks. To prepare our dataset (Section ~\ref{sec:datapreperation}), we applied the recently proposed configuration settings for GRU deep learning model by Karampatsis et al. \cite{karampatsis2020big}, which outperforms n-gram models on code completion and bug detection tasks. They used byte-pair encoding (BPE) technique to solve the unlimited vocabulary problem \cite{allamanis2013mining}.  This problem makes it infeasible to train LMs on large corpora. BPE is an algorithm originally designed for data compression, in which bytes that are not used in the data replace the most frequently occurring byte pairs or sequences \cite{gage1994new}. BPE starts by splitting all the words in characters. The initial vocabulary contains all the characters in the data set and a special end-of word symbol @@, and the corpus is split into characters plus @@. Then, it finds the most common pair of successive items in the corpus (initially characters, then tokens). This pair is merged in a new token which is added to the vocabulary; all occurrences of the pair are replaced with the new token. The process is repeated n times, which is called a merge operation (MO). 
We applied static settings with a large training set (50 epochs, 64 mini-batch size) and chose 10000 BPE MOs as it performs better than other BPE MOs such as 2000 and 5000. Static settings have been used to train a model on a fixed training corpus, and later evaluated on a separate test dataset. To train the LM, we first learned encoding by using the training set with the help of subword library \footnote{\url{https://github.com/rsennrich/subword-nmt}}. Then, we segmented the training, validation, and test sets using the learned encoding, and applied the MOs from BPE to merge the characters into subword units in the vocabulary.

\subsection{Generative Pretrained Transformer 2 (GPT-2)}
\label{sec:deepclonemodel}
OpenAI developed a large-scale unsupervised LM called GPT-2 (Generative Pretrained Transformer 2) \cite{radford2019language} to generate several sound sentences of realistic text by extending any given seed. GPT-2 is a large transformer-based LM with 1.5 billion parameters, trained on a dataset of 8 million web pages. GPT-2 is trained with a simple objective: predict the next word, given all of the previous words within some text. We focus on fine-tuning a  GPT-2 transformer \cite{radford2019language} pre-trained model for generating code clones, even though it has been trained on English language. We applied fine-tunning of a pre-trained model on IJaDataset - a Java language dataset - as there exists a large amount of overlapping vocabulary with English language. GPT-2 transformer has demonstrated impressive effectiveness of pre-trained LMs on various tasks including high quality text generation, question answering, reading comprehension, summarization, and translation \cite{radford2019language}.

GPT-2 also has built in BPE tokenizer. We selected a small version of GPT2 (GPT2-117) as our base model, as it does not take too much time and resources to fine-tune, and is enough to evaluate our approach. The GPT2-117\cite{radford2019language} pre-trained model has vocabulary size of 50257, 117M parameters, 12-hidden layers, 768-hidden states, and 12-attention heads. We have fine-tuned our GPT-2 based model on the partition of a GPU-1080Ti cluster (276 CPU cores, 329728 CUDA cores, 5.9 TB memory)\footnote{\url{https://userinfo.surfsara.nl/}} for approximately 9 hours by using HuggingFace Transformer Library. In our experiment, we have performed training and evaluation with batch size per GPU of 1 for 5 epochs. We have used a learning rate of 5e-5 and the gradient accumulation steps (number of update steps to accumulate before performing a backward/update pass) as 5. Default values have been used for other hyper-parameters, as mentioned in the language modeling code\footnote{\url{https://github.com/huggingface/transformers/blob/master/examples/language-modeling}}. 

\begin{figure}
  \centering
  \includegraphics[width=9.5cm,height=8.5cm]{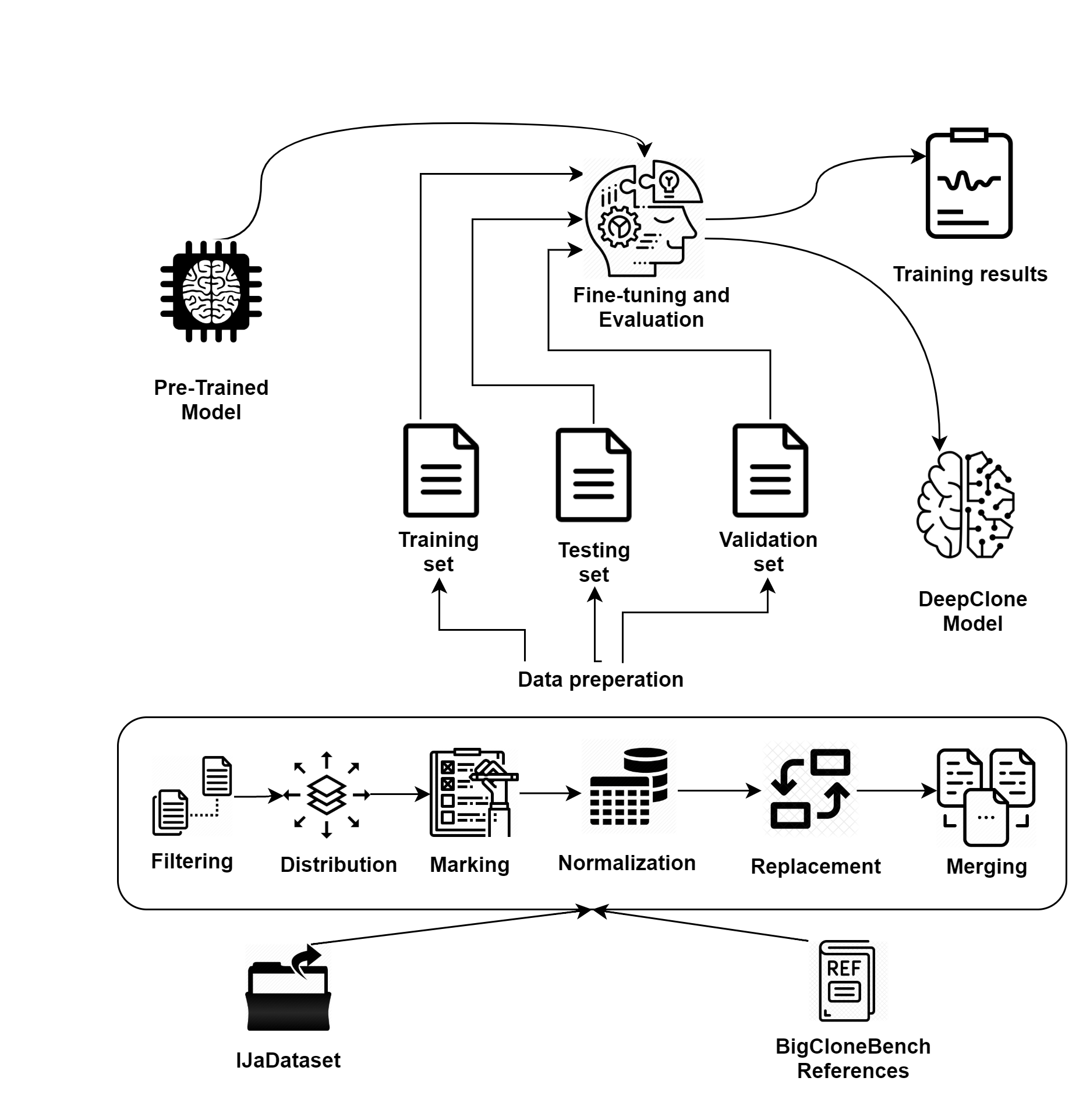}
  \caption{DeepClone training process\label{fig:deepmodeltraining}}
\end{figure}

\section{Comparative Evaluation: GRU vs GPT-2 Based Models}
We perform both intrinsic and extrinsic evaluations of GRU and GPT-2 based models to compare their performance. We calculate the perplexity scores (as done in related work \cite{zaremba2014recurrent,white2015toward}) to measure the quality of models (i.e.~intrinsic evaluation), which is an inverse of cross-entropy (as used in \cite{hellendoorn2017deep,karampatsis2020big}). Perplexity is a measurement of how well a given LM predicts sample data. It estimates the average number of code tokens to select from at each point in a sequence \cite{allamanis2013mining}. It is a natural evaluation metric for LMs, which represent a probability distribution over a subsequence or an entire dataset (Equation~\ref{eq:eq6}):

\begin{equation}
\label{eq:eq6}
P(L) = exp(-\frac{1}{M} \sum_{i}^{M} \log P(t_i|t_0 : t_{i-1}))
\end{equation}

\noindent
$P(t_i | t_0:t_{i-1})$ is the conditional probability assigned by the model to the token $t$ at index $i$. By applying $log$ of conditional probability, cross-entropy loss is calculated. $M$ refers to the length of tokens. Hence, perplexity is an exponentiation of the average cross entropy loss from each token $[0, M]$. We calculate the perplexity on the validation set (\textbf{P1}) and the testing set (\textbf{P2}) for GRU and GPT-2 based models, which clearly displays that the GPT-2 based model outperforms the other by a large margin (Table~\ref{tab:evaluationresults}). 

We further measure the performance of both models on specific tasks such as token prediction (i.e.~extrinsic evaluation). Given a number of code sequences as input, we collect the top 10 predictions from GRU and GPT-2 based models, and compute the top-k accuracy (the fraction of times the correct prediction appears in the top k predictions) for k $\in$ [1, 10]. Moreover, we measure the Mean Reciprocal Rank (MRR) scores of both language models (LM), which has been used by many researchers for evaluating code prediction such as \cite{karampatsis2020big,hellendoorn2017deep}. For each prediction done by the LM, we collect a ranked list of 10 predictions. For each of those lists, the reciprocal rank corresponds to the multiplicative inverse of the rank of the first correct answer. MRR in turn is the average of reciprocal ranks for all the input sequences used in the evaluation.

Table~\ref{tab:evaluationresults} shows the top-k accuracies as well as the MRR scores. Clearly, the results suggest that the GPT-2 based model performs more accurately compared to the GRU based model on pre-processed Java source code containing clone methods. The table also indicates that there is almost 77.808\% chance to get a correct token in the first option, and 94.999\% chance to have a correct output in the top-10 predicted outcomes for GPT-2 based model. To further quantify the accuracy of our models for token prediction task, we report an MRR score of 83\%, which indicates an excellent performance in evaluating a ranked list of predictions for GPT-2 based model. As GPT-2 based model gives us highest performance in terms of perplexity on validation set (\textbf{P1}) and test set (\textbf{P2}), MRR, and top-k accuracy, we continue with that model for the rest of paper, named it as DeepClone model for further evaluation.

\begin{table*}
\footnotesize{
\centering
\caption{Comparative evaluation results for GPT-2 and GRU models\label{tab:evaluationresults}}
\begin{tabular}{|l|r|r|l|r|r|r|r|}
\hline
\textbf{}                              & \multicolumn{2}{l|}{\textbf{Perplexities}}                                              & \multicolumn{5}{l|}{\textbf{Accuracies}}                                                                                                                                                          \\ \hline

\textbf{Model} & \textbf{Validation (P1)}& \textbf{Test (P2)} & \textbf{MRR} & \textbf{Top 1} & \textbf{Top 3} & \textbf{Top 5} & \textbf{Top 10} \\ \hline
\textbf{GPT-2}         & 2.145  & 2.146    & 84.329\%     & 77.808\%       & 90.040\%       & 92.766\%       & 94.999\%        \\ \hline
\textbf{GRU}           & 13.92   & 13.86     & 73.507\%     & 66.948\%       & 79.0715\%      & 82.02\%        & 84.787\%        \\ \hline
\end{tabular}
}
\end{table*}

\section{Further Evaluation of DeepClone Model}
In this section we describe further evaluations of DeepClone on additional aspects of the model.

\subsubsection{Training Evaluation} 
\begin{figure}
\centering
\begin{subfigure}{.29\textwidth}
  \centering
\includegraphics[scale=0.2]{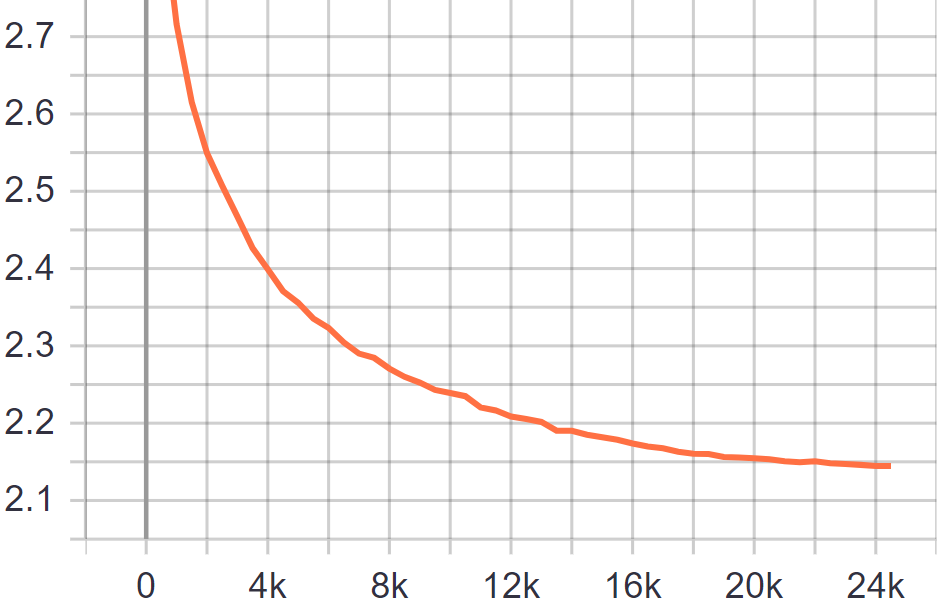}
  \caption{Perplexity over the validation dataset\label{fig:perplexity}}
\end{subfigure}%
\begin{subfigure}{.32\textwidth}
  \centering
\includegraphics[scale=0.2]{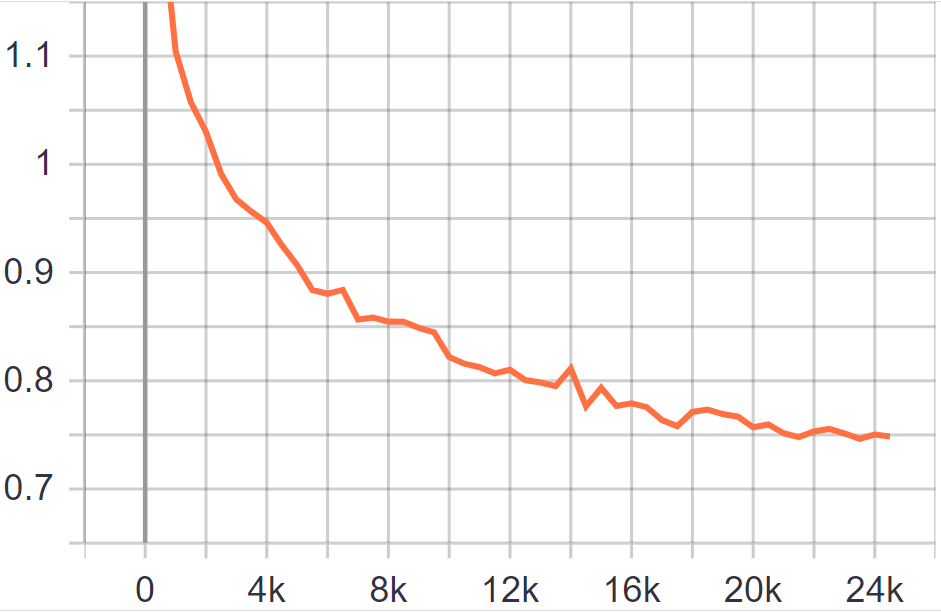}
  \caption{Training average losses\label{fig:loss}}
\end{subfigure}
\begin{subfigure}{.32\textwidth}
  \centering
\includegraphics[scale=0.2]{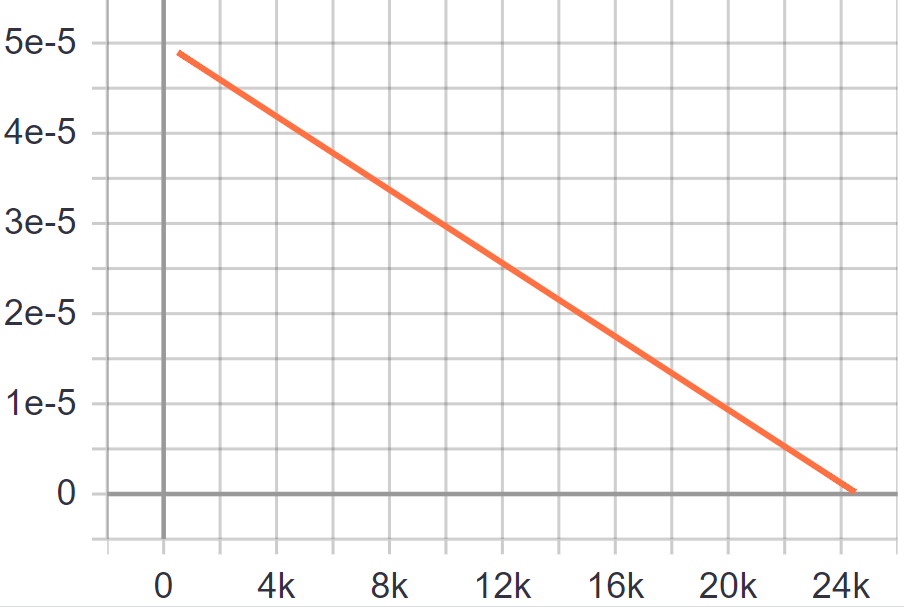}
  \caption{Convergence of the learning rate\label{fig:learningrate}}
\end{subfigure}
\caption{Training graphs\label{fig:training-graphs}}
\end{figure}

At each checkpoint (the 500th logging step) of the training steps, we evaluate DeepClone model performance by calculating the perplexity on the validation set. Figure ~\ref{fig:perplexity} describes the variations in perplexity on the validation set after each checkpoint. We observe that we achieve lowest perplexity \textbf{P1} (2.145) at step 24500. Figure~\ref{fig:learningrate} displays the convergence of the learning rate after each checkpoint. Learning rate helps in determining how quickly a neural network model learns a problem by adjusting the weights of a network with respect to the value of loss function. Another measure called loss function calculates a model error, which identifies how well a model predicts the expected outcome for any data point in the training set. GPT-2 uses cross-entropy loss function, which is to measure the performance of a LM whose output is a probability value between 0 and 1. Figure ~\ref{fig:loss} displays a convergence of training losses after each checkpoint, which indicates how well the model behaves after each checkpoint of optimization. The loss value is finally minimized to 0.75 at step 24500, which is a sign of a well optimized deep learning model. Figure ~\ref{fig:deepmodeltraining} describes the training process of the GPT-2 based model, which mentions the steps described in Section ~\ref{sec:bigclonebench} that are used to perform the fine-tuning of our model. All these numbers imply a successful training and an accurate model. We have published our training results online\footnote{\url{https://tensorboard.dev/experiment/tk1XqDi8RMqtrMjmVyQ9Sg}}.

\subsubsection{The Effect of Using Clone Markers}
\label{sec:clonemarkervsnonclonemarker}
Besides the overall perplexity on the testing dataset (\textbf{P2}), we re-calculate the perplexity using the testing dataset but without the clone method markers (i.e.~\big \langle soc\big\rangle~and \big \langle eoc\big\rangle). The motivation for this additional measurement is as follows. Hindle et al.~\cite{hindle2016naturalness} observed that due to the repetitive nature of the code, there exist predictable statistical properties, which n-gram language models can capture and leverage for software engineering tasks. The sign of a good model is that it can capture the patterns in the dataset very well, which is particularly important for the task of clone method prediction. In Table ~\ref{tab:evaluationresults}, we can see an increase (3.6\%) when comparing the original perplexity score of 2.146 (\textbf{P2}) and the perplexity on the test dataset without clone markers of 2.182 (\textbf{P3} see Table ~\ref{tbl:perplexities}), showing that DeepClone performs better when the code has marked clone methods. 

\subsubsection{Evaluation per Clone Method}
In order to determine which clone method snippets are more predictable compared to the others, we calculate average perplexity score ($\overline{PPL}$) for each functionality type (see Table \href{https://www.win.tue.nl/~mhammad/deepclone.html}{\textcolor{red}{A5}}). 
We first extract the code snippet for each type of clone method for our testing dataset, and calculate the perplexity score. The scores, as depicted in Table  \href{https://www.win.tue.nl/~mhammad/deepclone.html}{\textcolor{red}{A5}}, 
indicate how likely these can be predicted by DeepClone model. We also analyze several factors which can affect the perplexity of clone methods. BigCloneBench contains syntactic similarity scores for each clone method pair on the basis of tokens, calculated by using a line-based metric after normalization. We calculated the mean ($\mu$) and variance ($\sigma^{2}$) values to determine the overall syntactic similarity of all the clone methods per each type of functionality, as listed in Table  \href{https://www.win.tue.nl/~mhammad/deepclone.html}{\textcolor{red}{A5}}.

We observe that the perplexity scores vary according to the syntactic similarity between clone methods, as well as the number of clone method snippets in the training set. From the results, we can see that the "Test palindrome" type of clone method (number 44), which is used to test if a string is a palindrome, has the lowest perplexity score. It's thus well predicted by DeepClone. We attribute this to the high mean syntactic similarity (0.903$\pm$0.040) among those types of clone methods, and the relatively small number of snippets (133) used in training. Too few number of snippets in the training may lead to (a) high perplexities and low predictability e.g.~for "GCD" (number 26) to find the greatest common denominator and (b) no evaluation performed for "Decompress zip archive" clone method (number 5). Note that factors beside syntactical similarity and number of clone methods in the training set can also affect the perplexity score. In BigCloneBench, there are many false positive clone methods and other non-clone code snippets, which may be syntactically similar to true positive clone methods. Other factors such as clone types and hyper-parameters for GPT-2 are left to be explored in future work. 

\subsubsection{Non-Clone Methods vs Clone Methods}
Allamanis \cite{allamanis2019adverse} noticed that a language model achieves low perplexity scores for code duplication, and high perplexity score for less duplicated code. In order to observe that difference, we calculated the perplexity scores for all the clone method snippets and non-clone method snippets in the testing dataset. We extracted clone method snippets by tracing the tokens, which come inclusively between \big \langle soc \big \rangle  and \big \langle eoc \big \rangle  tokens. All other snippets were considered to be a part of non-cloned code. We then calculate the perplexity for each snippet. Finally, we take an average of the perplexities for both type of code snippets. Table~\ref{tbl:perplexities}, \textbf{P4} represents the average perplexity score for the clone method snippets, and \textbf{P5} represents the average perplexity of the non-cloned method snippets. We performed one-tailed Wilcoxon rank sum test to statistically compare \textbf{P4} with \textbf{P5}, which indicates that P4 is indeed less than P5 (\textit{p}<0.001). This shows that DeepClone correctly predicts clone method snippets much better than non-cloned snippets in general. 

\subsubsection{Performance on Other Datasets}
To evaluate the performance of DeepClone on another Java dataset, we use Allamanis et al.'s corpus \cite{allamanis2013mining} that contains over 14 thousand popular Java projects from Github. For base-lining, we focus only on 38 test projects that have been used in previous studies \cite{hellendoorn2017deep,karampatsis2020big}. We follow the same steps for dataset preparation as mentioned in Section ~\ref{sec:datapreperation},i.e., normalization, replacement, and merging. The dataset does not contain clone markers as no corresponding clones reference benchmark is available. As the main  purpose of clone markers is to help in predicting clone methods, so it will not severely affect the results of predicting next tokens, as also noticed in Section ~\ref{sec:clonemarkervsnonclonemarker}. On this dataset, we achieve a perplexity of 2.996 (\textbf{P6}) equivalent to 1.097 as cross-entropy. We further calculate other accuracy measures like MRR (81.523\%),  top-1 (74.416\%), top-3 (87.613\%), top-5 (90.704\%), and top-10 (93.152\%). These results (see Table~\ref{tbl:comparison}) outperform the static settings of previous studies \cite{hellendoorn2017deep,karampatsis2020big}. This indicates that DeepClone model is perfectly  fine-tuned with GPT-2 over Java corpus in general, and it contains an excessive amount of overlapping vocabulary with Allamanis et al.'s~\cite{allamanis2013mining} selected corpus.

\begin{table}
\centering
\caption{Performance of Allamanis et al.'s~\cite{allamanis2013mining} dataset on different models \label{tbl:comparison}}
\scriptsize
\begin{tabular}{|l|l|l|l|}
\hline
\textbf{Model} & \textbf{Settings}              & \textbf{Cross-Entropy} & \textbf{MRR} \\ \hline
LSTM/300 \cite{hellendoorn2017deep}       & Static                         & 3.22                  & 66.1\%       \\ \hline
LSTM/650\cite{hellendoorn2017deep}         & Static                         & 3.03                  & 67.9\%       \\ \hline
BPE NLM (512)  \cite{karampatsis2020big}            & BPE 10000, Static, Small train & 4.77                   & 63.75\%      \\ \hline
DeepClone      &     Section ~\ref{sec:deepclonemodel}                           & 1.097                 & 81.523\%     \\ \hline
\end{tabular}
\end{table}

\section{Clone Method Prediction}
In this section, we demonstrate how clone methods can be predicted from DeepClone model on the basis of user context. Furthermore, we measure its various aspects like rapid development and quality. 

\subsection{Experimental Design}
For predicting a clone method based on the user context, there exist several text generation methods such as beam search \cite{vijayakumar2018diverse} and nucleus sampling \cite{holtzman2019curious}. All these methods have a specific decoding strategy to shape the probability distribution of LM with higher probabilities assigned to higher quality texts. We selected nucleus sampling as it is claimed to be best the strategy for generating large amount of high quality text, comparable to human written text \cite{holtzman2019curious}. By using a fine-tuned model and nucleus sampling, we can expect a coherent set of code tokens for clone method prediction. Holtzman et al. \cite{holtzman2019curious} have also achieved coherent text generation results with similar settings.

We performed a small scale (100 context queries) experiment to predict next token subsequences by choosing different subsequence sizes of 10, 20, 30, 50, and 100 tokens. Among these, subsequences with size 20 gave us the best results in terms of top-k accuracy and MRR. We extracted subsequences of 20 tokens from the testing dataset, and moved the sliding window one step ahead to obtain further subsequences. From these we randomly selected 735 subsequences containing a total of 14,700 tokens, in which \big \langle soc\big \rangle~token is a part of each subsequence, indicating a start of clone method. We passed these one by one to DeepClone model, and kept on predicting new tokens with nucleus sampling (threshold value 0.95) until the meta-token \big \langle eoc\big \rangle~(i.e.~end of clone) appeared. We used the text generation script\footnote{\url{https://github.com/huggingface/transformers/blob/master/examples/text-generation/run\_generation.py}} of HuggingFace Transformer Library in this case. Note that certain parameters, e.g.~the number of subsequences and size of tokens per subsequence are chosen to perform a preliminary evaluation, to be fine-tuned and optimized in a follow-up study. The focus of this paper is to demonstrate the feasibility of our methodology for predicting clone methods. We have mentioned examples from our results in Tables (\href{https://www.win.tue.nl/~mhammad/deepclone.html}{\textcolor{red}{A1}}, \href{https://www.win.tue.nl/~mhammad/deepclone.html}{\textcolor{red}{A2}}, \href{https://www.win.tue.nl/~mhammad/deepclone.html}{\textcolor{red}{A3}}). To make the outputs readable in the tables, we have formatted the code by using the online tool\footnote{\url{https://www.tutorialspoint.com/online\_java\_formatter.htm}} along with little manual editing, which we plan to automate in future.

\subsection{Evaluation}

\subsubsection{Rapid Development}
In this experiment, we successfully generated 92,926 tokens associated with clone methods. Given the 735 cases, this amounts to an average of $\sim$126 tokens per case. As a comparison, other approaches\cite{vijayakumar2018diverse,holtzman2019curious} traditionally employ a threshold-based strategy of generating a certain number of code tokens up to a maximum threshold value of t. Note that t is typically a low value, e.g.~1 for simple token prediction, and 5-20 for the popular Deep TabNine auto-completer. Nevertheless, we can see that DeepClone, with the clone marking strategy, is able to outperform threshold-based strategies even with an extremely generous configuration of t = 50 or even 100. Furthermore, threshold-based strategies may not generate a set of code tokens for a complete method in a single pass, as the length of complete method varies. Marking the clone method regions in the dataset helps DeepClone to generate complete methods in a single pass. We conclude DeepClone model not only helps developers to code rapidly, but also provides a coherent set of code tokens for a complete method.

\subsubsection{Quality}
We measured the quality of DeepClone output by using ROUGE (Recall-Oriented Understudy for Gisting Evaluation) \cite{lin2004rouge}. It is designed to compare an automatically generated summary or translation against a set of reference summaries (typically human-generated). In our context, it helps us to automatically determine the quality of original DeepClone output by comparing it with the ground truth. ROUGE doesn't try to assess how fluent the clone method is. It only tries to assess the adequacy by simply counting how many n-grams in the DeepClone output match the ones in the ground truth. As ROUGE is based only on token overlap, it can determine if the same general concepts are discussed between an automatic and a reference summary, but it cannot determine if the result is coherent or the clone method is semantically correct. High-order n-gram ROUGE measures try to judge fluency to some degree.

We calculated precision (P), recall (R), and F-measure (F) of ROUGE-1 ROUGE-2, and ROUGE-L between DeepClone output and ground truth. ROUGE-1 refers to the overlap of unigrams between some reference output and the output to be evaluated. ROUGE-2, in turn, checks for bigrams instead of unigrams. The reason one would use ROUGE-1 over or in conjunction with ROUGE-2 (or other finer granularity ROUGE measures), is to also indicate fluency as part of the evaluation. The intuition is that the prediction is more fluent if it more closely follows the word orderings of the reference snippet. Finally, ROUGE-L measures longest matching sequence of tokens between machine generated text/code and human produced one by using longest common subsequence (LCS). LCS has a distinguishing advantage in evaluation: it captures in-sequence (i.e.~sentence level flow and word order) matches rather than strict consecutive matches. DeepClone predicted clone method can be extremely long, capturing all tokens in the retrieved clone methods, but many of these tokens may be useless, making it unnecessarily verbose. This is where precision comes into play. It measures what portion of the DeepClone output is in fact relevant and desirable to be kept with respect to the reference output.

\begin{equation}
\label{eq:eqprecision}
\text{Precision}=\frac{\text{\# of overlapping tokens}}{\text{total tokens in the predicted output}}
\end{equation}
Recall in the context of ROUGE measures what portion of the reference output was successfully captured by the DeepClone output.

\begin{equation}
\label{eq:eqrecall}
\text{Recall}=\frac{\text{\# of overlapping tokens}}{\text{total \# tokens in reference snippet}}
\end{equation}

We also report the F-measure which provides a single score that balances both the concerns of precision and recall.
\abovedisplayskip=0.05pt
\begin{equation}
\label{eq:eqpfscore}\text{F-Measure} = 2*\frac{\text{Precision * Recall}}{\text{Precision + Recall}}
\end{equation}
\belowdisplayskip=0.05pt

We have measured different ROUGE scores, i.e.~ROUGE-1, ROUGE-2, and ROUGE-L, to evaluate the similarity (and the quality to a certain extent) of the DeepClone output to ground truth. In this step, we extract only the tokens between \big \langle soc\big \rangle~and \big \langle eoc\big \rangle (inclusive) from the DeepClone output and ground truth (see Tables \href{https://www.win.tue.nl/~mhammad/deepclone.html}{\textcolor{red}{A1}}, \href{https://www.win.tue.nl/~mhammad/deepclone.html}{\textcolor{red}{A2}}, and \href{https://www.win.tue.nl/~mhammad/deepclone.html}{\textcolor{red}{A3}}). We observe quite reasonable scores for ROUGE-1 and ROUGE-L against ROUGE-2 (Table~\ref{tbl:results_gtpo}). This depicts that the DeepClone output contains reasonably well overlap of uni-grams and longest sequence matches of tokens with the ground truth compared to bi-grams.
 
In our qualitative investigation, we experienced two different scenarios based on the input context. The first one is when the context contains the method name. It is straightforward for the neural language technique to generate the predicted clone method following the given method name and current context. Table  \href{https://www.win.tue.nl/~mhammad/deepclone.html}{\textcolor{red}{A1}} gives an example of this scenario, where "transpose" method name is mentioned in the context and our approach predicts the clone method, whose functionality type matches the ground truth. The second scenario is based on the context that does not contain a method name. This can have two different output sub-scenarios. The first one is when the functionality type of the ground truth do not match. As we see in Table  \href{https://www.win.tue.nl/~mhammad/deepclone.html}{\textcolor{red}{A3}}, 
the context does not have the full signature of the clone method. This makes the generated output by DeepClone using nucleus sampling deviate from the functionality type of the ground truth. Ground truth belongs to "copy file" functionality, while DeepClone output belongs to "delete directory", which eventually leads to low and largely deviating ROUGE scores between the DeepClone output and the ground truth (see Table ~\ref{tbl:results_gtpo} and the example in Table \href{https://www.win.tue.nl/~mhammad/deepclone.html}{\textcolor{red}{A3}}. 
These clone methods may or may not fulfil the desired goal of the user. So, it might be useful to guide the users to include the clone method name in the context for better results. The other output sub-scenario is when we manage to successfully generate DeepClone output whose functionality type matches with the ground truth. In Table \href{https://www.win.tue.nl/~mhammad/deepclone.html}{\textcolor{red}{A2}}, 
"copy file" method name is not mentioned in the context, but the functionality type of the DeepClone output matches with the ground truth. We notice that the total number of "copy file" clone methods used in DeepClone training are 2,454, which allows nucleus sampling to generate DeepClone output closer to ground truth in example \href{https://www.win.tue.nl/~mhammad/deepclone.html}{\textcolor{red}{A2}}. 
Overall, we believe our approach yields good results and can assist the developers by correctly predicting clone methods in different scenarios.

\begin{table*}
\footnotesize{

\parbox{.4\linewidth}{
\hfill

\caption{Final Distribution of BigCloneBench Dataset\label{tab:experimentalcorpus}}
\begin{tabular}{|l|r|r|r|}
\hline
                    & 
                    \multicolumn{1}{l|}{\textbf{Files}} & \multicolumn{1}{l|}{\textbf{Clone Methods}} & \multicolumn{1}{l|}{\textbf{Tokens}} 
                    \\ \hline
\textbf{Training}   & 9,606                               & 11,991                                      & 16,933,894                          
\\ \hline
\textbf{Validation} & 1,208                               & 1,499                                       & 2,130,360                            
\\ \hline
\textbf{Testing}    & 1,234                               & 1,502                                       & 2,235,982                            
\\ \hline
\textbf{Total}      & 12,048                              & 14,992                                      & 21,300,236                          
\\ \hline
\end{tabular}
\vspace{0.7cm}
\noindent
\centering
\caption{Perplexities\label{tbl:perplexities}}
\begin{tabular}{|l|l|l|l|}
\hline %
\textbf{P3} & \textbf{P4} & \textbf{P5} & \textbf{P6}\\ \hline 
2.182       & 2.410       & 2.767 & 2.996\\ \hline  
\end{tabular}
}
\hfill
\parbox{.4\linewidth}{
\centering

\caption{Empirical evaluation results between DeepClone output and ground truth\label{tbl:results_gtpo}}
\begin{tabular}{|l|l|}
\hline
\textbf{ROUGE-1}   &                   \\ \hline
\textbf{Precision} & 0.667 $\pm$ 0.192           \\ \hline
\textbf{Recall}    & 0.559   $\pm$ 0.226      \\ \hline
\textbf{F-measure}  & 0.56 $\pm$     0.185    \\ \hline
\textbf{ROUGE-2}   &                   \\ \hline
\textbf{Precision} & 0.479 $\pm$    0.217     \\ \hline
\textbf{Recall}    & 0.398 $\pm$ 0.218         \\ \hline
\textbf{F-measure}  & 0.4 $\pm$  0.202       \\ \hline
\textbf{ROUGE-L}   &                   \\ \hline
\textbf{Precision} & 0.652  $\pm$  0.165      \\ \hline
\textbf{Recall}    & 0.586  $\pm$  0.183      \\ \hline
\textbf{F-measure}  & 0.599  $\pm$  0.153      \\ \hline
\end{tabular}

}
}
\end{table*}

\section{Discussion}
Our approach leads to promising results. The performance metrics in the training (learning rate approaching 0, minimized loss) and validation (perplexity of 2.145) phases all indicate a fine-tuned model. The series of calculated perplexity scores allow us to conclude that DeepClone model can predict regularities successfully in terms of clone markers, including the code in general and the individual clone snippets in particular. The extrinsic evaluation reveals that we achieve high accuracy, notably 95\% in the top 10 suggestions, as well as larger number of tokens than a threshold-based strategy even with a generous threshold of 100. With a high quality and accurate model as the foundation, we next discuss the potential use cases to exploit our model, as well as the limitations to our work.

\subsubsection{Use Cases for DeepClone}
\label{sec:usecases}
DeepClone model can be utilized to assist developers in various scenarios. Some of these have already been mentioned above: predicting the next token (as typically done by many LMs) or the complete clone method body. The latter, while seemingly straightforward, can be enhanced with a more elaborate ranking and retrieval mechanism rather than simply generating the most likely sequence of tokens one after another. For that purpose, the additional information in BigCloneBench, including the exact clone method clusters (methods representing the same functionality), clone types, and so on can be exploited. Another use case might involve clone refactoring (and avoidance), by recommending extract method refactoring instead of predicting a complete clone method snippet. In combination with some additional code transformations, the clone methods can be converted to reusable assets (e.g.~in the form of libraries). The model can also be used to perform code search for common functionalities.

\subsubsection{Limitations and Threats to Validity} 
Our approach is the first step for code prediction that raises the granularity level to complete methods. However, we cannot expect exactly the same clone method being predicted or completed as the one used in training by DeepClone model. In prediction tasks, generating well-formed outputs is challenging, which is well-known in natural language generation \cite{hashimoto2018retrieve}. The desired output might be a variation of another, previously observed sample \cite{hashimoto2018retrieve}, due to the probabilistic nature of the LM; the space of possible clone methods that could be generated grows exponentially with the length of the clone methods. An extension of the current work would involve displaying the most similar cloned methods (as is) from the dataset to the user. 

Some limitations originate from the selected dataset. BigCloneBench only contains clone references of methods for the 43 common functionalities, i.e.~not all the clones are marked in the dataset. Although it is enough to validate our methodology, modeling all the clones might result in more interesting findings. Although BigCloneBench is a well-known dataset, it does not necessarily represent Java language source code entirely (a threat to external validity). 

In our study, we relied on the HuggingFace transformer implementation of GPT-2 to train and evaluate DeepClone model. While GPT-2 is a reliable architecture used in many NLP experiments \cite{radford2019language}, HuggingFace transformer implementation is still an emerging project. However, our results and trends are aligned with those obtained in the field of NLP. Hence, we are positive that the results are reliable. As for the clone method prediction, we have only used nucleus sampling. Other techniques such as beam search can also be explored.
\section{Conclusion and Future Work}
In this work, we proposed DeepClone, a deep learning based cloned code language model. We have performed intrinsic and extrinsic evaluations to determine its performance in predicting clone methods. The extensive evaluation suggests that our approach significantly improves code prediction by exploiting deep learning and code clones. In future work, we plan to implement the potential use cases of this model (Section ~\ref{sec:usecases}). The proposed LM can be improved by hyper-parameter optimizations, as well as by better training (e.g.~on a larger dataset or larger pre-trained GPT-2 models). We also plan to investigate how to tackle different types and granularity levels of code clones such as simple clones, structural clones, file clones, and clones of other artifact types such as models\cite{hammad2020systematic,babur2019clone}.

\section*{Acknowledgment}
Dr. Sohaib Khan (CEO at \href{http://hazen.ai/}{Hazen.ai}) provided valuable feedback on the experimentation. \href{https://userinfo.surfsara.nl/}{SURFsara} provided credits for experiments. The project is partly funded by Prince Sultan University Faculty Research Fund.

\bibliographystyle{splncs04}
\bibliography{paper-references}

\end{document}